\documentstyle{article}

\newtheorem{theorem}{\sc Theorem}
\newtheorem{lemma}{\sc Lemma}
\newtheorem{coro}{\sc Corollary}
\newtheorem{nota}{\sc Notation}
\newtheorem{defin}{\sc Definition}
\newtheorem{cla}{\sc Claim}
\newtheorem{rem}{\sc Remark}
\newtheorem{ex}{\sc Example}
\newcommand{\K}{C}
\newcommand{\KP}{K}

\newenvironment{proof}{\par \sc Proof.\rm}{\hspace*{\fill}$\Box$\vspace{1ex}}
\newenvironment{example}{\begin{ex}}{\hspace*{\fill}$\Diamond$\end{ex}}
\newenvironment{claim}{\begin{cla}}{\end{cla}}
\newenvironment{remark}{\begin{rem}}{\end{rem}}
\newenvironment{corollary}{\begin{coro}}{\end{coro}}
\newenvironment{definition}{\begin{defin}}{\end{defin}}

\title{A New Approach to
Formal Language Theory by
Kolmogorov Complexity\thanks{A preliminary version of part
of this work was presented at the %
\it 16th
International Colloquium on Automata, Languages,
and Programming%
\rm , Stresa, Italy, July 1989.}}

\author{
Ming Li\thanks{Supported
in part by
National Science Foundation Grant DCR-8606366,
Office of Naval Research Grant
N00014-85-k-0445, Army Research Office Grant DAAL03-86-K-0171,
and by NSERC operating grants OGP-0036747 and OGP-046506.
Part of the work was performed while
he was with the
Department of Computer Science, York University,
North York, Ontario, Canada.
Address: Computer Science Department,
University of Waterloo,
Waterloo, Ontario, Canada N2L 3G1. Email: mli@math.waterloo.edu}\\
University of Waterloo\\
\and
Paul M.B. Vit\'{a}nyi\thanks{Partially
supported by NSERC
International Scientific Exchange Award ISE0046203, and
by NWO through NFI Project ALADDIN under Contract
number NF 62-376.
Address: CWI,
Kruislaan 413, 1098 SJ Amsterdam, The Netherlands. Email: paulv@cwi.nl}\\
CWI and Universiteit van Amsterdam \\}

\begin{document}
\bibliographystyle{plain}
\maketitle
\begin{abstract}
We present a new approach to formal language theory using Kolmogorov
complexity. The main results presented here are
an alternative for pumping
lemma(s), a new characterization for regular languages, and
a new method to separate deterministic context-free languages
and nondeterministic context-free languages.
The use of the new `incompressibility arguments' is illustrated by
many examples.
The approach is also successful at the high end
of the Chomsky hierarchy since one can quantify nonrecursiveness
in terms of Kolmogorov complexity. 
(This is a preliminary uncorrected version. The final
version is the one published in
 {\em SIAM J. Comput.}, 24:2(1995), 398-410.)
\end{abstract}

\section{Introduction}
It is feasible to reconstruct
parts of formal language theory using
algorithmic information theory (Kolmogorov complexity).
We provide theorems on how to
use Kolmogorov complexity as a concrete and powerful
tool. We do not just want to introduce
fancy mathematics; our goal is
to help our readers do a large part
of formal language theory in the most essential,
usually easiest, sometimes even obvious ways.
In this paper it is only important to us to demonstrate
that the application of Kolmogorov complexity
in the targeted area
is not restricted to trivialities.
The proofs of the theorems in this paper may not be easy.
However, the theorems are of the type that are
used as a tool. Once derived, our theorems
are easy to apply.
\subsection{Prelude}

The first application of Kolmogorov complexity in
the theory of computation was in \cite{Pa79,PSS81}.
By re-doing proofs of known results, it was shown that
static, descriptional (program size) complexity of a %
\it single %
\rm random
string can be used to obtain lower
bounds on dynamic, computational (running time) complexity.
None of the inventors of Kolmogorov complexity 
originally had these applications
in mind.
Recently, Kolmogorov complexity 
has been applied extensively
to solve classic open problems
of sometimes two decades standing, \cite{Ma84,LiVi88a,JiLi93,JiVi93}.
For more examples see the textbook
\cite{LiVi90}. 

The secret of Kolmogorov complexity's success
in dynamic, computational lower bound proofs rests on a simple
fact: the overwhelming majority of strings
has hardly any computable regularities. 
We call such a string `Kolmogorov random' or `incompressible'.
A Kolmogorov random string cannot be (effectively) compressed.
Incompressibility is a noneffective property:
no individual string, 
except finitely many, can be proved incompressible.

Recall that a traditional lower bound proof by 
counting usually involves %
\it all %
\rm inputs
of certain length. One shows that a certain lower bound
has to hold for %
\it some `typical' %
\rm input. Since an individual
typical input is %
\it hard %
\rm (sometimes impossible)
to find, the proof has to involve all the
inputs. Now we understand that a typical input of each length
can be constructed via an
incompressible string. However,
only finitely many
individual strings can be 
effectively proved to be incompressible. 
No wonder the old counting
arguments had to involve all inputs. 
In a proof using the new
`incompressibility method',
one uses an individual incompressible string that is known to
{\it exist} even though it cannot be constructed.
Then one shows that if the assumed lower time bound
would not hold, then this string could be compressed, 
and hence it would not be incompressible.

\subsection{Outline of the Paper}

The incompressibility argument above also works
for formal languages and automata theory proper.
Assume the basic notions treated in a textbook like \cite{HoUl79}.

The first result is a powerful alternative to pumping lemmas for regular
languages. It is well known that not all nonregular languages
can be shown to be nonregular by the usual $uvw$-pumping lemma.
There is a plethora of pumping lemmas to show nonregularity,
like the `marked pumping lemma', and so on. In fact, it seems that
many example nonregular languages require their own special
purpose pumping lemmas.
Comparatively recently,  \cite{Ja78,StWe82,EPR81}, 
exhaustive 
pumping lemmas that characterize the regular
languages have been obtained.

These pumping lemmas are complicated and complicated to use.
The last reference uses Ramsey theory. In contrast, using Kolmogorov
complexity we give a new characterization of
the regular languages that simply makes our intuition
of `finite state'ness of these languages rigorous
and is easy to apply. Being a characterization
it works for all non-regular languages. We give several
examples of its application, 
some of which were quite difficult using pumping lemmas.

To prove that a certain context-free language (cfl) is not 
deterministic context-free (dcfl) has
required laborious ad hoc proofs, \cite{HoUl79},
or cumbersome--to--state and also difficult--to--apply
pumping lemmas or iteration theorems \cite{Ha78,Yu89}.
We give necessary (Kolmogorov complexity) conditions for dcfl, 
that are very easy to apply. We test the new method on several examples
in $\rm{cfl} -{\rm dcfl}$, which were hard
to handle before. In certain respects the KC-DCFL Lemma 
may be more powerful than the related lemmas and
theorems mentioned above.
On the high end of
the Chomsky hierarchy we present, for completeness, a known
characterization of recursive languages, and a necessary condition
for recursively enumerable languages.

\section{Kolmogorov Complexity}
From now on,
let $x$ denote both the natural number and the
$x$th binary string in the sequence $0, 1, 00, 01, 10, 11, 000, \ldots $
That is, the representation `$3$' corresponds both to the
natural number $3$ and to the binary string
$00$. This way we obtain a natural bijection between the
nonnegative integers ${\cal N}$ and the finite binary
strings $\{0,1\}^*$. Numerically, the binary string $x_{n-1} \ldots x_0$ 
corresponds to the integer
\begin{equation}\label{eq.numstring}
  2^n - 1 + \sum_{i=0}^{n-1} x_i 2^i . 
\end{equation}

We use notation $ l(x) $ to denote the
\it length \rm%
(number of bits) of a binary string $x$.
If $x$ is not a finite binary string but another finite object like
a finite automaton, a recursive function, or a natural number, then we
use $l(x)$ to denote the length of its standard
binary description.
Let $ \langle \cdot , \cdot  \rangle : {\cal N} \times {\cal N}
\rightarrow {\cal N}$ be a
standard recursive, invertible, one-one encoding
of pairs of natural numbers in natural numbers. This idea 
can be iterated to obtain a pairing from triples of natural numbers
with natural numbers
$\langle x,y,z \rangle = \langle x, \langle y,z \rangle \rangle$,
and so on.

Any of the usual definitions of Kolmogorov complexity in
\cite{Ko65,PSS81,LiVi90}
will do for the sequel.
We are interested in the shortest effective description
of a finite object $x$. To fix thoughts,
consider the problem of describing a string $x $ over
0's and 1's. Let $T_1 , T_2 , \ldots$ be the standard enumeration
of Turing machines. Since $T_i$ computes a partial recursive
function $\phi_i : {\cal N} \rightarrow {\cal N}$ 
we obtain the standard enumeration
$\phi_1 , \phi_2 , \ldots$ of partial recursive functions.
We denote $\phi (\langle x,y \rangle )$ as $\phi (x,y)$.
Any partial recursive function $\phi$ from strings over
0's and 1's to such strings, together with a string $p $, the
\it program %
\rm for $\phi$ to compute $x$, such that
$\phi( p ) = x $, is a description of $x$. 
It is useful to generalize this idea to the
conditional version: $\phi ( p, y ) = x$
such that $p$ is a program for $\phi$ to compute $x$,
given a binary string $y$ for free.
Then the %
\it descriptional \rm%
complexity $\K_{\phi}$ of $x $, %
\it relative \rm%
to $\phi$ and $y $, is defined by
$$
\K_{\phi} ( x | y ) = \min  \{  l(p) : p \in  \{ 0,1 \}^*, \phi (p, y ) = x \} ,
$$
or $\infty$ if no such $p$ exists. 

For a $universal$ partial recursive
function $\phi_0$, computed by the universal Turing machine $U$, we know that,
for each partial recursive function $\phi$, there is a constant 
$c_{\phi}$ such that 
for all strings $x,y$, we have $\phi_0 (i,x,y)= \phi (x,y)$. Hence,
$\K_{{\phi}_0 }( x | y )  \leq  \K_{\phi} ( x | y ) + c_{\phi}$.
We fix a reference universal function $\phi_0$
and define
the %
\it conditional Kolmogorov complexity %
\rm of $x$ given $y$
as $\K(x| y ) = \K_{\phi_0} (x| y )$.
\footnote{Similarly, we define the complexity of the $x$th partial
recursive function $\phi$ conditional to the $y$th
partial recursive function $\psi$ by $\K(\phi|\psi) = \K(x|y)$.}

The %
\it unconditional Kolmogorov
complexity %
\rm of $x $ is $\K( x ) = \K( x | \epsilon )$%
\rm ,
where $\epsilon$ denotes
the empty string ($l( \epsilon )=0$).

Since there is a Turing machine that
just copies its input to its output we have $\K(x| y)  \leq   l(x)  + O(1)$,
for each $x$ and $y$.
Since there are $2^n$ binary
strings of length $n$, but only $2^n -1$ possible shorter descriptions
$d$, it follows that $\K(x)  \geq   l(x) $ for some binary string $x$ 
of each length.
We call such strings $incompressible$ or $random$. It also
follows that, for any length $n$ and any binary string $y$,
there is a binary string $x$ of length $n$ such that
$\K(x|y)  \geq   l(x) $. Considering $\K$ as an integer function,
using the obvious one-one correspondence between finite binary
words and nonnegative integers, it can be shown that
$\K(x) \rightarrow \infty$ for $x \rightarrow \infty$.
Finally, $\K(x, y)$ denotes $\K( \langle  x, y \rangle )$.

\begin{example}[Self-Delimiting Strings]
\rm
A
\it prefix code %
\rm is a mapping from finite binary code words
to source words, such that no code word is a proper prefix
of any other code word. We define a particular prefix code.

For each binary source word $x=x_1 \ldots x_n$,  
define the code word  $\bar x$ by
\[ \bar x=1^{l(x)}0x .\]
Define
\[ x' = \overline{l(x)} x. \]
The string $x'$ is called the %
\it self-delimiting \rm%
code of $x$.

Set $x=01011$. Then, $l(x)=5$, which corresponds to binary string
`10', and $\overline{l(x)} = 11010$. Therefore,
$x'= 1101001011$ is the self-delimiting code of `01011'.

The self-delimiting code of a positive integer $x$ requires
$ l(x)  + 2 \log  l(x) +1 $ bits. It is
easy to verify that $l(x) = \lfloor \log (x+1) \rfloor$.
All logarithms are base 2 unless otherwise noted. For convenience,
we simply denote the length $l( x )$ 
of a natural number $x$ by `$\log x$'.
\end{example}

\begin{example}[Substrings of incompressible strings]
\rm
Is a substring of an incompressible string also incompressible?
A string $x = uvw$ can be specified by a short description
for $v$ of length $\K(v)$, a description of $ l(u) $,
and the literal description of $uw$. Moreover, we need
information to tell these three items apart. Such
information can be provided by prefixing each item
with a self-delimiting description of its length.
Together this
takes $\K(v) + l(uw) + O( \log  l(x) )$ bits. Hence,
$$
\K(x)  \leq  \K(v)+ O( \log  l(x) )+l(uw)  ,
$$
Thus, if we choose $x$ incompressible, $\K(x)  \geq   l(x) $, then we obtain
$$
\K(v)  \geq  l(v) - O( \log  l(x) )  .
$$
It can be shown that this is optimal --- a substring of
an incompressible string of length $n$
can be compressible by an $O( \log n)$ additional term. This
conforms to a fact we know from probability
theory: every random string
of length $n$ is expected to contain a run of about
$\log n$ consecutive zeros (or ones). Such a substring has
complexity $O( \log \log n)$.
\end{example}

\section{Regular Sets and Finite Automata}
\begin{definition}
\rm
Let $\Sigma$ be a finite nonempty alphabet, and let $Q$
be a (possibly infinite) nonempty set of states.
A {\em transition function} is a function
$\delta: \Sigma \times Q  \rightarrow  Q $.
We extend $\delta$ to $\delta'$ on $\Sigma^*$
by $\delta' ( \epsilon ,q)=q$ and
$$
\delta'  (a_1 \ldots a_n,q) =
\delta (a_n , \delta' (a_1 \ldots a_{n-1},q)) .
$$
Clearly, if $\delta' $ is not $1-1$,
then the automaton `forgets'
because some $x$ and $y$ from $\Sigma^* $
drive $\delta'$ into the same
memory state. An {\em automaton} $A$ is a quintuple
$(\Sigma,Q,\delta,q_0,q_f)$ where everything is as above
and $q_0,q_f \in Q$ are distinguished {\em initial state}
and {\em final state}, respectively.
We call $A$ a {\em finite automaton} (fa) if $Q$ is finite.
\end{definition}

We denote
`indistinguishability' of a pair of histories
$x,y \in  \Sigma^* $ by $x \sim y$, defined as
$\delta' (x,q_0 )=\delta' (y, q_0 )$. `Indistinguishability'
of strings is reflexive, symmetric, transitive,
and right-invariant ($\delta'(xz,q_0) = \delta'(yz, q_0)$ for all $z$).
Thus, `indistinguishability'
is a right-invariant equivalence relation on $\Sigma^* $.
It is a simple matter
to ascertain this formally.
\begin{definition}\index{regular language}
\rm
The language accepted by automaton $A$ as above is the
set $L=\{x: \delta'(x,q_0)=q_f\}$. A {\em regular language}
is a language accepted by a finite automaton.
\end{definition}

It is a straightforward exercise
to verify from the definitions the
following fact (which will be used later).

\begin{theorem}[Myhill, Nerode]\index{Theorem!Myhill-Nerode}
\label{theo.myhill-nerode}
The following statements
about $L  \subseteq \Sigma^* $ are equivalent.

{\rm (i)} $L \subseteq \Sigma^* $ is accepted by some finite automaton.

{\rm (ii)} $L$ is the union of equivalence classes
of a right-invariant equivalence relation of finite index on $\Sigma^* $.

{\rm (iii)} For all $x,y \in \Sigma^* $ define right-invariant equivalence
$x \sim y$ by:
for all $z \in \Sigma^* $ we have $xz \in L$ iff $yz \in L$.
Then the number of $\sim$-equivalence classes is finite.
\end{theorem}

Subsequently, closure of finite automaton languages
under complement, union, and intersection
follow by simple construction of the
appropriate $\delta$ functions
from given ones. Details can be found in any textbook
on the subject like \cite{HoUl79}.
The clumsy pumping lemma approach can now be
replaced by the Kolmogorov formulation below.

\subsection{Kolmogorov Complexity Replacement for the Pumping Lemma}

An important part of formal language theory is deriving
a hierarchy of language families. The main division is
the Chomsky hierarchy, with regular languages, context-free
languages, context-sensitive languages and recursively enumerable
languages. The common way to prove that certain languages
are not regular is by using `pumping' lemmas,
for instance, the $uvw$-lemma. However, these lemmas are
quite difficult to state and cumbersome to prove or use. In contrast,
below we show how to replace such arguments by simple, intuitive
and yet rigorous, Kolmogorov complexity arguments. 

Regular languages coincide with the languages accepted by
finite automata. This invites a straightforward
application of Kolmogorov complexity.
Let us give an example. 
We prove that $ \{ 0^k 1^k : k \geq 1 \} $
is not regular. If it were, then the state $q $ of a particular accepting fa
after processing $0^k $, together with the fa,
is, up to a constant, a description
of $k$. 
Namely, by running $A$, initialized in state $q$,
on input consisting of only 1's, the first time $A$
enters an accepting state is after precisely $k$
consecutive 1's.
The size of the description
of $A$ and $q$ is bounded by a constant, say $c$,
which is independent of $k$.
Altogether, it follows that
$\K(k)  \leq  c +O(1)$. But choosing $k$ with $\K(k) \geq \log k$
we obtain a contradiction for all large enough $k$.
Hence, since the fa has a fixed finite number of states,
there is a fixed finite number that bounds
the Kolmogorov complexity of each natural number: contradiction.
We generalize this observation as follows.

\begin{definition}
\rm
Let $\Sigma$ be a finite nonempty alphabet, and let 
$\phi : {\cal N} \rightarrow  \Sigma^* $
be a total recursive function. Then $\phi$ enumerates
(possibly a proper subset of) $ \Sigma^* $ in order $\phi (1), \phi (2), \ldots $
We call such an order %
\it effective%
\rm , and $\phi$ an
\it enumerator%
\rm .
\end{definition}
The %
\it lexicographical order %
\rm is the effective order such that
all words in $ \Sigma^* $ are ordered first according to length, 
and then lexicographically
within the group of each length. Another example is $\phi$ such that
$\phi (i) = p_i$, the standard binary representation of the $i$th
prime, is an effective order in $ \{ 0, 1 \}^*$. In this
case $\phi$ does not enumerate all of $ \Sigma^* $.
Let $L  \subseteq   \Sigma^* $. Define $L_x =  \{ y: xy  \in  L \} $.

\begin{lemma}[KC-Regularity]\label{kc-regularity}
Let $L  \subseteq   \Sigma^* $ be regular, and let $\phi$ an enumerator in $ \Sigma^* $.
Then there exists a constant $c$ depending only on $L$ and $\phi$,
such that for each $x$, if $y$ is the $n$th string
enumerated in (or in the complement of) 
$L_x$,
then $\K(y)  \leq  \K(n) + c$. 
\end{lemma}
\begin{proof}
Let $L$ be a regular language.
The $n$th string $y$ such that $xy  \in  L$ for some $x$ can be described
by
\begin{itemize}
\item
this discussion, and
a description of the fa that accepts $L$;
\item
a description of $\phi$; and
\item
the state of the fa after processing $x$, and the number $n$. 
\end{itemize}\par\noindent
The statement ``(or in the complement of)'' follows,
since regular languages
are closed under complementation.
\end{proof}

As an application of the KC-Regularity Lemma we prove that
$ \{ 1^p : p$ %
\it is prime$ \} $ %
\rm is not regular. 
Consider the string $xy=1^p$ with $p$ the
$(k+1)$th prime. Set $x=1^{p'}$, with $p'$ the $k$th prime.
Then $y=1^{p-p'}$, and $y$ is the lexicographical first
element in $L_x$. Hence, by Lemma~\ref{kc-regularity},
$\K(p-p')=O(1)$. But the difference between two
consecutive primes grows unbounded. Since there
are only $O(1)$ descriptions of length $O(1)$,
we have a contradiction.
We give some more examples from the well-known textbook
of Hopcroft and Ullman that are marked * as difficult there:
\begin{example}[Exercise 3.1(h)* in \cite{HoUl79}]
\rm
Show $L=  \{  xx^R w :  x,w  \in   \{ 0,1 \}^* -  \{  \epsilon  \} $\} 
is not regular. Set $x = (01)^m$, where
$\K(m)  \geq  \log m$. Then, the lexicographically first
word in $L_x$ is $y$ with $y = (10)^m 0$.
But, $\K(y) = \Omega ( \log m )$, contradicting
the KC-Regularity Lemma.
\end{example}
\begin{example}
\rm
Prove that $L=\{0^i 1^j : i \neq j \}$ is not regular. Set $x=0^m$,
where $\K(m) \geq \log m$. Then, the lexicographically first
word {\em not} in $L_x \bigcap \{1\}^*$ is $y = 1^m$. 
But, $\K(y) = \Omega ( \log m )$, contradicting
the KC-Regularity Lemma.
\end{example}

\begin{example}[Exercise 3.6* in 
\cite{HoUl79}]
\rm
Prove that $L= \{ 0^i 1^j : \gcd (i,j)=1$\} is not regular.
Set $x=0^{(p-1)!} 1$, where $p>3$ is a prime, $l(p)=n$ and
$\K(p) \geq \log n - \log \log n$. Then the
lexicographically first word in $L_x$ is $1^{p-1}$,
contradicting the KC-regularity lemma.
\end{example}
\begin{example}[Section 2.2, Exercises 11-15, \cite{Ha78}]
\rm
Prove that $ \{ p:p$ is the standard binary representation of a
prime $\}$
is not regular. Suppose the contrary, and $p_i$ denotes
the $i$th prime, $i \geq 1$.
Consider the least binary $p_m =uv $ ($=u2^{l(v)}+v$),
with $u=\Pi_{{i < k}} p_i$
and $v$ not in $\{0\}^* \{1\}$. Such a prime $p_m$ exists since
each interval $[n,n+n^{11/20} ]$ of the natural numbers contains
a prime, \cite{HeIw79}.

Considering $p_m$ now as an integer,
$p_m =2^{l(v)} \Pi_{i < k} p_i +v$. Since
integer $v > 1$ and $v$ is not divided by any prime
less than $p_k$ (because $p_m$ is prime),
the binary length $l(v) \geq l(p_k )$.
Because $p_k$ goes to infinity with $k$, the value
$\K(v) \geq \K(l(v))$ also goes to infinity with $k$.
But since $v$ is the lexicographical first suffix,
with integer $v > 1$,
such that $uv \in L$, we have
$\K(v)=O(1)$ by the KC-Regularity Lemma,
which is a contradiction.
\end{example}

\subsection{Kolmogorov Complexity Characterization of Regular Languages}

While the pumping lemmas are not precise
enough (except for the difficult 
construction in \cite{EPR81}) to characterize the regular languages, 
with Kolmogorov complexity this is easy. In fact, the 
KC-Regularity Lemma
is a direct corollary of the characterization below.
The theorem is not only a device to
show that some nonregular languages are nonregular,
as are the common pumping lemmas,
but it is a 
\it characterization %
\rm of the regular sets. Consequently, it
determines whether or not a given language is regular,
just like the Myhill-Nerode Theorem. The usual
characterizations of regular languages seem to be practically useful
to show regularity. The need for pumping lemmas
stems from the fact that characterizations tend to
be very hard to use
to show nonregularity.
In contrast, the KC-characterization
is practicable for both purposes, as evidenced by the examples.

\begin{definition}
\rm
Let $\Sigma$ be a nonempty finite alphabet,
and let $y_i$ be the $i$th element of $\Sigma^*$ in
lexicographic order, $i \geq 1$.
For $L \subseteq \Sigma^*$ and
$x  \in  \Sigma^* $, let $\chi = \chi_1 \chi_2  \ldots $
be the %
\it characteristic
sequence\index{characteristic sequence!of a language|bold}
\rm of $L_x = \{y:xy \in L \}$,
defined by $\chi_i  =  1$ %
\rm if $xy_i   \in  L$, and $\chi_i =0$ otherwise. We denote
$\chi_1 \ldots  \chi_n$ by $\chi_{1:n}$.
\end{definition}

\begin{theorem}[Regular KC-Characterization]
Let $L \subseteq \Sigma^*$, and assume the notation above. 
The following statements are equivalent. 

{\rm (i)} $L$ is regular.

{\rm (ii)} There is a constant $c_L $ depending 
only on $L$, such that
for all $x  \in   \Sigma^* $, for all $n$,
$\K( \chi_{1:n} | n)  \leq  c_L $.

{\rm (iii)} There is a constant $c_L $ depending 
only on $L$, such that
for all $x  \in   \Sigma^* $, for all $n$,
$\K( \chi_{1:n} )  \leq  \K(n) + c_L$.

{\rm (iv)} There is a constant $c_L $ depending 
only on $L$, such that
for all $x  \in   \Sigma^* $, for all $n$,
$\K( \chi_{1:n} )  \leq  \log n + c_L$.
\end{theorem}

\begin{proof}
\rm
(i) $\rightarrow$ (ii): by similar proof as the KC-Regularity Lemma.

\rm
(ii) $\rightarrow$ (iii): obvious.

\rm
(iii) $\rightarrow$ (iv): obvious.

(iv) $\rightarrow$ (i):

\begin{claim}
\rm
\label{claim.finite}
For each constant $c$ there are only finitely many
one-way infinite binary strings $\omega $ such
that, for all $n$, 
$\K( \omega_{1:n} )  \leq  \log n + c$.
\end{claim}
\begin{proof}
\rm
The claim is a weaker version of Theorem 6
in \cite{Ch76}.
It turns out that the weaker version 
admits a simpler proof. 
To make the treatment self-contained we present
this new proof in the Appendix.
\end{proof}

By (iv) and the claim, there are only finitely many distinct
$\chi$'s associated with the $x$'s in $ \Sigma^* $.
Define the right-invariant
equivalence relation $\sim$  by $x \sim x'$ if
$\chi = \chi '$. This relation induces a partition of $ \Sigma^* $
in equivalence classes $[x]= \{ y: y  \sim  x \} $.
Since there is a one-one correspondence between the $[x]$'s
and the $\chi$'s, and
there are only finitely many distinct $\chi$'s,
there are also only finitely many $[x]$'s, which implies
that $L$ is regular by the Myhill-Nerode theorem. 
\end{proof}

\begin{remark}
\rm
The KC-regularity Lemma may be viewed as a corollary of
the Theorem. If $L$ is regular, then 
clearly $L_x$ is regular, and it follows immediately
that there are only finitely many associated $\chi$'s,
and each can be specified in at most $c$ bits, where 
$c$ is a constant depending
only on $L$ (and enumerator $\phi$).
If $y$ is, say, the $n$th
string in $L_x$, then we can specify $y$ as the string corresponding
to the $n$th `1' in $\chi$, using only $\K(n)  +  O(1)$ bits to specify
$y$. Hence $\K(y)   \leq   \K(n) + O(1)$. Without loss of generality,
we need to assume
that the  $n$th string
enumerated in $L_x$ in the KC-regularity Lemma is the string
corresponding to the $n$th `1' in $\chi$ by the enumeration in the Theorem,
or that there is a recursive mapping between the two.
\end{remark}
\begin{remark}
\rm
If $L$ is nonregular, then
there are infinitely many $x  \in   \Sigma^* $
with distinct equivalence classes $[x]$, each of which has
its own distinct associated characteristic sequence $\chi$.
It is easy to see, for each automaton (finite or infinite),
for each $\chi$ associated with an equivalence class $[x]$ we have 
$$
\K ( \chi_{1:n} | n) \rightarrow  \inf \{ \K(y): y  \in  [x] \}  + O(1),
$$
for $n \rightarrow \infty$. The difference between finite
and infinite automata is precisely expressed in
the fact that only in the first case does there exist
an a priori constant which bounds the lefthand term
for all $\chi$. 
\end{remark}

We show how to prove positive results
with the KC-Characterization Theorem. (Examples of negative results
were given in the preceding section.)

\begin{example}
\rm
Prove that $L =  \Sigma^* $ is regular. There
exists a constant $c$, such that for each $x$ the
associated characteristic sequence is $\chi = 1,1, \ldots $,
with $\K( \chi_{1:n} | n)  \leq  c$.
Therefore, $L$ is regular by the KC-Characterization Theorem.
\end{example}
\begin{example}
\rm
Prove that $L =  \{ x: x \mbox{ the number of `1's in $x$ is odd} \}$ 
is regular. 
Obviously, there exists a constant $c$
such that for each $x$ we have $\K( \chi_{1:n} )  \leq  \K( n ) + c$.
Therefore, $L$ is regular by the KC-Characterization Theorem.
\end{example}

\section{Deterministic Context-free Languages}

We present a Kolmogorov complexity based criterion
to show that certain languages are not dcfl.
In particular, it can be used
to demonstrate the existence of witness languages
in the difference of the family of context-free languages
(cfls) and deterministic context-free languages (dcfls). 
Languages in this difference are the most difficult
to identify; other non-dcfl are also non-cfl
and in those cases we can often 
use the pumping lemma for context-free languages. The new method
compares favorably with other known related techniques 
(mentioned in the Introduction) by being
simpler, easier to apply, and apparently more powerful
(because it works on a superset of examples).
Yet, our primary goal is to demonstrate the usefulness
of Kolmogorov complexity in this matter.

A language is a dcfl iff it is accepted by a deterministic
pushdown automaton (dpda). 

Intuitively, the lemma below tries to capture the following.
Suppose a dpda accepts $L=\{0^n 1^n 2^n : n \geq 1 \}$.
Then the dpda needs to first
store a representation of the all-0 part, and then retrieve it to check
against the all-1 part.
But after that check, it seems inevitable that it has
discarded the relevant information about $n$,
and cannot use this information again to check against
the all-2 part. That is,
the complexity of the all-2 part should be $\K(n)=O(1)$,
which yields a contradiction for large $n$.
\begin{definition}\label{def.recursive.string}
\rm
A one-way infinite string $\omega = \omega_1 \omega_2 \ldots$ over $\Sigma$ is
{\it recursive}
if there is a total recursive function $f: {\cal N} \rightarrow \Sigma$
such that $\omega_i = f(i)$ for all $i \geq 1$.
\end{definition}

\begin{lemma}[KC-DCFL]\label{lemma.KC-DCFL}
Let $L \subseteq \Sigma^* $ be recognized by a deterministic pushdown
machine $M$
and let $c$ be a constant.
Let $\omega = \omega_1 \omega_2 \ldots$
be a recursive sequence over $\Sigma$ which can be
described in $c$ bits.
Let $x,y \in \Sigma^*$ with $\K(x,y) < c$ and let
$\zeta = \ldots \zeta_2 \zeta_1$ be a (reversed) recursive
sequence over $\Sigma$ of the form $ \ldots yyx$.
Let $n,m \in {\cal N}$ and $w \in \Sigma^*$ be such that Items 
{\rm (i)} to {\rm (iii)} below
are satisfied.

{\rm (i)} 
For each $i$ ($1 \leq i \leq n$),
given $M$'s state and pushdown store contents after processing
input $\zeta_m \ldots \zeta_1 \omega_1 \ldots \omega_i$,
a description of $\omega$,
and an additional description of at
most $c$ bits,
we can reconstruct $n$
by running $M$ and observing only
acceptance or rejection.

{\rm (ii)}
Given $M$'s state and pushdown store contents after processing
input $\zeta_m \ldots \zeta_1 \omega_1 \ldots \omega_n$, we can reconstruct 
$w$ from an additional description of at most $c$ bits.

{\rm (iii)} $K(\omega_1 \ldots \omega_n)  \geq   2 \log \log  
m $.

\noindent
Then there is a constant $c'$ depending only on $L$ and $c$
such that $\K(w)  \leq  c'$.
\end{lemma}

\begin{proof}
\rm 
Let $L$ be accepted by $M$ with input head $h_r$.
Assume $m,n,w$ satisfy the conditions in the statement
of the lemma. For convenience we write
\[ u = \zeta_m \ldots \zeta_1, \hspace{2em}
v = \omega_1 \ldots \omega_n . \]
For each input $z \in \Sigma^*$, we denote with $c(z)$
the pushdown store contents at the time $h_r$ has read all
of $z$, and moves to the right adjacent input symbol.
Consider the computation of $M$ on input $uv$
from the time when $h_r$ reaches the end
of $u$. There are two cases:

\vspace{0.1in}
\bf
Case 1.
\rm
There is a constant $c_1$ such that
for infinitely many pairs $m,n$ satisfying the statement of the lemma
if $h_r$ continues and reaches
the end of $v$, then all of the original $c(u)$
has been popped except at most the bottom $c_1$ bits.

That is, 
machine $M$
decreases its pushdown store
from size $l(c(u))$ to size $c_1$ during the processing of $v$.
The first time this occurs, let $v'$ be the processed initial segment of $v$,
and $v''$ the unprocessed suffix (so that $v=v'v''$)
and let $M$ be in state $q$. We can describe $w$ by the following
items.\footnote{
Since we need to glue different binary items in the encoding together,
in a way so that we can effectively separate them again,
like $\langle x,y \rangle = x' y$, we count $\K(x) + 2 \log \K(x)+1$ bits for
a self-delimited encoding $x'=1^{l(l(x))} 0 l(x) x$ of $x$ . We only need
to give self-delimiting forms for all but one constituent description item.}
\begin{itemize}
\item
A self-delimiting
description of $M$ (including $\Sigma$) and this discussion in $O(1)$ bits.
\item
A self-delimiting description of $\omega$ in $(1+\epsilon)c$ bits.
\item
A description of $c(uv')$ and $q$ in $c_1 \log |\Sigma| + O(1)$ bits.
\item
The `additional description' mentioned in Item (i)
of the statement of the lemma in self-delimiting format,
using at most $(1+\epsilon)c$ bits. Denote it by $p$.
\item
The `additional' description mentioned  in Item (ii)
of the statement of the lemma in self-delimiting format,
using at most $(1+\epsilon)c$ bits. Denote it by $r$.
\end{itemize}
By Item (i) in the statement of the lemma we
can reconstruct $v''$ from $M$ in state $q$ and with pushdown store
contents $c(uv')$, and $\omega$, using description $p$.
Subsequently, starting $M$ in state $q$
with pushdown store contents $c(uv')$, we process $v''$.
At the end of the computation
we have obtained $M$'s state and pushdown store
contents after processing $uv$. According to Item (ii)
in the statement of the lemma, together with description $r$
we can now reconstruct $w$. Since $\K(w)$ is at most the length
of this description, 
\[ \K(w) \leq  4c+c_1 \log |\Sigma| + O(1) . \]
Setting $c' := 4c + c_1 \log |\Sigma| +O(1)$ satisfies the lemma.

\vspace{0.1in}
{\bf Case 2.}
By way of contradiction, assume that Case 1 does not hold.
That is, for each constant $c_1$ all but finitely many pairs $m,n$ satisfying
the conditions in the lemma cause
$M$ not to decrease its stack height below $c_1$ 
during the processing
of the $v$ part of input $uv$. 

Fix some constant $c_1$.
Set $m,n$ so that they satisfy the statement of the lemma,
and to be as long as required to validate the argument below.
Choose $u'$ as a suffix of $yy \ldots y x$ with $l(u') > 2^m$ and
\begin{equation}\label{def.uprime}
\K(l(u' ))   <   \log \log  m.
\end{equation}
That is, $l(u')$ is much larger than $l(u)$ ($=m$) and much more regular.
A moment's reflection learns that we can always choose such a $u'$.

\begin{claim}\label{claim.cycle}
\rm
For large enough $m$ there exists a $u'$ as above, such that
$M$ starts in the same state and
accesses the same top $l(c(u))-c_1$ elements of its stack during
the processing of the $v$ parts of both inputs $uv$ and $u'v$.
\end{claim}

\begin{proof}
\rm
By assumption, $M$ does not read below
the bottom $c_1$ symbols of $c(u)$ while processing the $v$ part
of input $uv$.

We argue that one can choose $u'$
such that the top segment of $c(u' )$
is precisely the same as the
top segment of $c(u)$ above the bottom $c_1$ symbols,
for large enough $ l(u) $, $ l(u') $.

To see this we examine the initial computation of $M$ on $u$.
Since $M$ is deterministic, it must either cycle through
a sequence of pushdown store contents, or increase its pushdown store
with repetitions on long enough $u$ (and $u'$).
Namely, let a triple $(q,i,s)$ mean 
that $M$ is in state $q$, has top pushdown store symbol
$s$, and $h_r$ is at $i$th bit of some $y$. Consider only
the triples $(q,i,s)$ at the steps where
$M$ will never go below the
current top pushdown store level again 
while reading $u$. (That is, $s$ will not be
popped before going into $v$.) There are precisely
$ l(c(u))$ such triples. Because the input is repetitious
and $M$ is deterministic, some triple must start to repeat within
a constant number of steps and with a
constant interval (in height of $M$'s pushdown store)
after $M$ starts reading $y$'s. It is easy to
show that within a repeating interval 
only a constant number of $y$'s are
read. 

The pushdown store does not cycle
through an a priori bounded set of pushdown store contents,
since this would mean that there
is a constant $c_1$ such that the processing by $M$
of any suffix of $yy \ldots y x$ 
does not increase the stack height above $c_1$. This situation
reduces to Case 1 with $v = \epsilon$.

Therefore, the pushdown store contents grows repetitiously and unboundedly.
Since the repeating
cycle starts in the pushdown store after a constant number of symbols,
and its size is constant in number of $y$'s, we can adjust
$u'$ so that $M$ starts in the same state
and reads the same top segments of $c(u)$ and $c(u' )$
in the $v$ parts
of its computations on $uv$ and $u' v$. 
This proves the claim. 
\end{proof}

The following items form a description from which we
can reconstruct $v$. 
\begin{itemize}
\item
This discussion and a description of $M$ in $O(1)$ bits.
\item
A self-delimiting description of the 
recursive sequence $\omega$ of which $v$ is
an initial segment in $(1+ \epsilon)c$ bits.
\item
A self-delimiting description of the pair $\langle x,y \rangle$ 
in $(1+ \epsilon)c$ bits.
\item
A self-delimiting description of $l(u')$ in 
$(1+ \epsilon )  \K(l(u'))$ bits.
\item
A program $p$ to reconstruct $v$
given $\omega$ and $M$'s state and pushdown 
store contents after processing $u$.
By Item (i)
of the statement of the lemma,
$l(p) \leq c$. Therefore, a self-delimiting description of $p$ 
takes at most $(1+\epsilon)c$ bits.
\end{itemize}
The following procedure reconstructs $v$ from this information.
Using the description of $M$ and $u'$ we construct the state  $q_{u'}$
and pushdown store contents $c(u')$ of $M$ after processing $u'$.
By Claim~\ref{claim.cycle}, the state $q_u$ of $M$ after processing
$u$ satisfies $q_u=q_{u'}$ and the top $l(c(u))-c_1$ elements
of $c(u)$ and $c(u')$ are the same. Run $M$ on input $\omega$
starting in state
$q_{u'}$ and with stack contents $c(u')$. By assumption,
no more than $l(c(u))-c_1$ elements of
$c(u')$ get popped before we have processed $\omega_1 \ldots \omega_n$.
By just looking at the consecutive states of $M$ in this computation,
and using program $p$,
we can find $n$ according to Item (i) in the statement of the lemma.
To reconstruct $v$ requires by definition at least $\K(v)$ bits.
Therefore,
\begin{eqnarray*}
 \K(v) & \leq & (1 + \epsilon )\K(l(u')+4c +O(1) \\
& \leq & (1 + \epsilon)  \log \log m + 4c + O(1) ,
\end{eqnarray*}
where the last inequality follows by Equation~\ref{def.uprime}.
But this contradicts Item (iii) in the statement of the lemma
for large enough $m$.
\end{proof}

Items (i) through (iii) in the KC-DCFL Lemma can be
considerably weakened,
but the presented version gives
the essential idea and power: it suffices for many examples.
A more restricted, but easier, version is the following.

\begin{corollary}\label{cor.kc-dcfl}
Let $L  \subseteq   \Sigma^* $ %
be a dcfl and let $c$ be a constant.
Let $x$ and $y$ be fixed finite words over $\Sigma$ and
let $\omega$ be a recursive sequence over $\Sigma$.
Let $u$ be a suffix of $yy \ldots y x$,
let $v$ be a prefix of $\omega$,
and let $w \in \Sigma^*$ such that:

{\rm (i)} $v$ can be described in $c$ bits given
$L_u$ in lexicographical order;

{\rm (ii)} $w$ can be described in $c$ bits given
 $L_{uv}$ in lexicographical order; and

{\rm (iii)} $\K(v)  \geq   2 \log \log  l(u) $.

\noindent
Then there is a constant $c'$ depending only on $L,c,x,y,\omega$
such that $\K(w)  \leq  c'$.
\end{corollary}

All the following context-free languages were proved
to be not dcfl only with great effort before,
\cite{HoUl79,Ha78,Yu89}. Our new proofs are 
more direct and intuitive. Basically,
if $v$ is the first word in $L_u$,
then processing the $v$ part of input $uv$ must have already used up
the information of $u$. But if there is not much information
left on the pushdown store, then the first word $w$ in $L_{uv}$
cannot have high Kolmogorov complexity.

\begin{example}[Exercise 10.5 (a)** in  \cite{HoUl79}]
\rm Prove $L =  \{ x: x = x^R , x  \in   \{ 0, 1 \}^*  \} $ %
\rm is
not dcfl. Suppose the contrary. Set $u=0^n 1$ and $v=0^n$, $\K(n) \geq \log n$,
satisfying Item (iii) of the lemma.
Since $v$ is lexicographically 
the first word in $L_u$, Item (i) of the lemma
is satisfied.
The lexicographically
first nonempty word in $L_{uv}$ is
$10^n$, and so we can set $w=10^n$ satisfying Item (ii) of the lemma.
But now we have
$\K(w) = \Omega ( \log n)$, contradicting
the KC-DCFL Lemma and its Corollary.

Approximately the same proof shows that 
the context-free language
$ \{ xx^R : x  \in   \Sigma^*  \} $ and
the context-sensitive language
$ \{ xx : x  \in   \Sigma^*  \} $ are not deterministic context-free languages.
\end{example}

\begin{example}[Exercise 10.5 (b)** in \cite{HoUl79}, 
Example 1 in~\cite{Yu89}]
\rm Prove
$ \{ 0^n 1^m :  m=n, 2n \} $ is not dcfl. 
Suppose the contrary.
Let $u=0^n$ and $v=1^n$,
where $\K(n)  \geq   \log n$.
Then $v$ is the lexicographically first word in $L_u$.
The lexicographically first nonempty
word in $L_{uv}$ is $1^n$.
Set $w = 1^n$, and $\K(w) = \Omega ( \log n)$, 
contradicting the KC-DCFL Lemma and its Corollary.
\end{example}
\begin{example}[Example 2 in~\cite{Yu89}]
\rm
Prove $L =  \{ xy:  l(x)  =  l(y) , y$ contains a `1', 
$x, y  \in   \{ 0, 1 \}^* \} $ %
\rm is not dcfl. 
Suppose the contrary. Set $u = 0^n 1$ where $l(u)$ is even.
Then $v= 0^{n+1}$ is lexicographically the first even length word
not in $L_u$. With $\K(n) \geq \log n$,
this satisfies Items (i) and (iii) of the lemma.
Choosing $w=10^{2n+3}$, the lexicographically first 
even length word not in $L_{uv}$ starting with a $`1'$,
satisfies Item (ii). But
$\K(w) = \Omega ( \log n)$, which contradicts the KC-DCFL Lemma
and its Corollary.
\end{example}

\begin{example}
\rm Prove
$L =  \{ 0^i 1^j 2^k : i,j,k \geq 0,  i=j \mbox{ or } j=k \} $
is not dcfl. Suppose the contrary.
Let $u = 0^n $ and $v = 1^n$ where $\K(n)  \geq  \log n$, satisfying
item (iii) of the lemma.
Then, $v$ is lexicographically the first word
in $L_u$, satisfying Item (i).
The lexicographic first word
in $L_{uv} \cap \{1\}\{2\}^*$ is $12^{n+1}$. Therefore,
we can set $w = 12^{n+1}$ and satisfy Item (ii). Then
$\K(w)  =  \Omega ( \log n)$, contradicting
the KC-DCFL Lemma and its Corollary.
\end{example}

\begin{example}[Pattern-Matching]
\rm
The KC-DCFL Lemma and its Corollary can be used trickily.
We prove $\{x\#yx^R z: x,y,z \in \{0,1\}^*\}$ is not dcfl.
Suppose the contrary. Let $u=1^n \#$, and $v=1^{n-1}0$ where
$\K(n) \geq \log n$, satisfying Item (iii) of the lemma.
Since $v'=1^n$ is the lexicographically first word
in $L_u$, the choice of $v$ satisfies Item (i) of the lemma.
(We can reconstruct $v$ from $v'$ by flipping
the last bit of $v'$ from 1 to 0.)
Then $w=1^n$ is lexicographically
the first word in $L_{uv}$, to satisfy Item (ii).
Since $\K(w)=  \Omega ( \log n)$,
this contradicts the KC-DCFL Lemma and its Corollary.
\end{example}

\section{Recursive, Recursively Enumerable, and Beyond}

It is immediately obvious how to characterize recursive languages
in terms of Kolmogorov complexity. If $L \subseteq  \Sigma^* $, and
$ \Sigma^* = \{ v_1 ,v_2 , \ldots  \} $ is effectively ordered, then
we define the characteristic sequence 
$\lambda = \lambda_1 , \lambda_2 ,  \ldots $ of $L$ by $\lambda_i =1$
if $v_i  \in L$ and $\lambda_i =0$ otherwise. 
In terms of the earlier developed
terminology, if $A$ is the automaton accepting $L$, then
$\lambda$ is the characteristic sequence associated with the
equivalence class $[ \epsilon ]$.  Recall Definition~\ref{def.recursive.string}
of a recursive sequence.
A set
$L \in \Sigma^*$ is recursive iff its characteristic
sequence $\lambda$ is a recursive sequence. It then
follows trivially from the definitions:

\begin{theorem}[Recursive KC Characterization]
A set $L \in  \Sigma^* $ is recursive, iff there exists a constant
$c_L$ (depending only on $L$) such that,
for all $n$, $\K( \lambda_{1:n} | n) < c_L$. 
\end{theorem}

$L$ is r.e. if the set $ \{ n:  \lambda_n =1 \} $ is r.e. In terms
of Kolmogorov complexity, the following theorem gives not
only a qualitative but even a quantitative difference between
recursive and r.e. languages. The following
theorem is due to Barzdin', \cite{Ba68,Lo69a}.

\begin{theorem}[KC-r.e.]

(i) If $L$ is r.e., then there is a constant $c_L$ (depending
only on $L$), such that for all $n$, $\K( \lambda_{1:n} |n)  \leq  \log n +c_L$.

(ii) There exists an r.e. set $L$ such that $\K( \lambda_{1:n} ) \geq \log n$,
for all $n$.
\end{theorem}

Note that, with $L$ as in Item (ii), the set
$ \Sigma^* -L$ (which is possibly non--r.e.)
also satisfies Item (i). Therefore, Item (i) is not
a Kolmogorov complexity characterization of the r.e. sets.

\begin{example}
\rm
Consider the standard enumeration of Turing machines. Define
$k=k_1  k_2   \ldots $ by $k_i =1$ if the $i$th Turing machine started
on its $i$th program halts ($\phi_i (i) < \infty$), 
and $k_i =0$ otherwise. Let $A$
be the language such that $k$ is its characteristic sequence.
Clearly, $A$ is an r.e. set.
In \cite{Ba68} it is shown that $\K(k_{1:n} ) \geq \log n$, for all $n$.
\end{example}

\begin{example}
\rm
Let $k$ be as in the previous example.
Define a one-way infinite binary sequence $h$ by
$$
h = k_1  0^2  k_2  0^{{2}^2}   \ldots 
k_i  0^{{2}^i}  k_{i+1}   \ldots 
$$
Then, $\K(h_{1:n} ) = O(\K(n))  +  \Theta ( \log \log  n)$.
Therefore, if $h$ is the characteristic sequence of a set $B$,
then $B$ is not recursive, but more `sparsely' nonrecursive
than is $A$.
\end{example}

\begin{example}
\rm
The probability that the optimal universal Turing machine $U$
halts on self-delimiting binary 
input $p$, randomly supplied by tosses of
a fair coin, is $\Omega$, $0 <  \Omega  <  1$. Let the
binary representation of $\Omega$ be $0. \Omega_1 \Omega_2 \ldots$
Let $\Sigma$ be
a finite nonempty alphabet, and $v_1 ,v_2 ,  \ldots $
an effective enumeration without repetitions of $ \Sigma^* $.
Define $L \subseteq  \Sigma^* $ such that $v_i  \in L$ iff $\Omega_i =1$.
It can be shown, see for example~\cite{LiVi90},
that the sequence $\Omega_1 , \Omega_2 , \ldots$ satisfies
\[ \K( \Omega_{1:n} |n)  \geq  n - \log n -2 \log \log n -O(1), \]
for all but finitely many $n$.

Hence neither $L$ nor $ \Sigma^* -L$ is r.e. It is not difficult to see
that $L \in  \Delta_2 -( \Sigma_1  \cup  \Pi_1 )$,
in the arithmetic hierarchy
(that is, $L$ is not recursively enumerable), \cite{La87,La89}.
\end{example}
\section{Questions for Future Research}

(1) It is not difficult to give a direct KC-analogue of the $uvwxy$ Pumping
Lemma (as Tao Jiang pointed out to us). Just like the Pumping
Lemma, this will show that $ \{ a^n b^n c^n : n  \geq  1 \} $,
$ \{ xx: x \in  \Sigma^*  \} $, $ \{ a^p : p$ is prime$ \} $, and so on,
are not cfl.
Clearly,
this hasn't yet captured the Kolmogorov complexity 
heart of cfl. More in general,
can we find
a CFL-KC-Characterization? 

(2) What about ambiguous context-free languages?

(3) What about context-sensitive languages and 
deterministic context-sensitive languages?

\section*{Appendix: Proof of Claim~\protect\ref{claim.finite}}

A {\em recursive real} is a real number whose binary
expansion is
recursive in the sense of Definition~\ref{def.recursive.string}.
The following result is demonstrated in \cite{Lo69b}
and attributed to A.R. Meyer.
For each constant $c$
there are only finitely many $\omega \in \{ 0,1 \}^{\infty}$ with
$\K( \omega_{1:n} |n) \leq c$ for all $n$.
Moreover, each such $\omega$ is a recursive real.

In \cite{Ch76} this is strengthened to
a version with $\K( \omega_{1:n} ) \leq \K( n ) +c$,
and strengthened again to
a version with $\K( \omega_{1:n} ) \leq \log n +c$.
Claim~\ref{claim.finite} is weaker than the latter version
by not requiring the
$\omega$'s to be recursive reals.
For completeness sake, we
present a new direct proof of Claim~\ref{claim.finite}
avoiding the notion of recursive reals.

Recall our convention of identifying integer
$x$ with the $x$th binary sequence in lexicographical order
of $\{0,1\}^*$ as in Equation~\ref{eq.numstring}.

\begin{proof}[of Claim~\ref{claim.finite}]
\rm
Let $c$ be a positive constant, and let
\begin{eqnarray}
\label{eq.An}
A_n & = & \{ x \in \{0,1\}^n : \K( x )  \leq  \log n + c \}, \\
\nonumber
A & = & \{ \omega \in \{0,1 \}^{\infty} : \forall_{n \in {\cal N}}
[ \K( \omega_{1:n} ) \leq \log n + c ] \} \:.
\end{eqnarray}

If the cardinality $d(A_n )$ of $A_n$
dips below a fixed constant $c'$,
for infinitely many $n$, then $c'$ is an upper bound on
$d(A)$. This is because it is an upper bound on
the cardinality of the set of prefixes of length $n$
of the elements in $A$, for {\em all} $n$.

Fix any $l \in {\cal N}$.
Choose a binary string $y$ of length $2l+c+1$ satisfying
\begin{equation}\label{random}
\K(y) \geq  2l+c+1.
\end{equation}
Choose $i$ maximum such that
for division of $y$ in $y=mn$ with $l(m)=i$
we have
\begin{equation}\label{eq.m}
m \leq d(A_n).
\end{equation}
(This holds at least for $i=0=m$.)
Define similarly a division $y=sr$
with $l(s)=i+1$.
By maximality of $i$, we have $s > d(A_{r})$.
From the easily proven $s \leq 2m+1$, it then follows that
\begin{equation}\label{bound}
d(A_{r}) \leq 2m.
\end{equation}

We prove $l(r) \geq l$.
Since by Equations~\ref{eq.m} and \ref{eq.An} we have
\[m \leq d(A_{n}) \leq 2^{c}n, \]
it follows that $l(m) \leq l(n)+c$.
Therefore,
\[ 2l+c+1 =l(y)= l(n)+l(m) \leq 2l(n)+c , \]
which implies that $l(n) > l$. Consequently,
$l(r) = l(n)-1 \geq l$.

We prove $d(A_{r}) = O(1)$.
By dovetailing the computations of the reference universal
Turing machine $U$ 
for all programs $p$ with $l(p) \leq \log n +c$,
we can enumerate all elements of $A_n$.
We can reconstruct $y$ from
the $m$th element, say $y_0$, of this enumeration.
Namely, from $y_0$ we reconstruct $n$ since $l(y_0)=n$,
and we obtain
$m$ by enumerating $A_n$ until $y_0$ is generated. By
concatenation we obtain $y=mn$. Therefore,

\begin{equation}\label{(3)}
\K(y) \leq \K(y_0)+O(1) \leq \log n + c + O(1) .
\end{equation}
From Equation~\ref{random} we have
\begin{equation}\label{(4)}
\K(y) \geq \log n + \log m .
\end{equation}
Combining Equations~\ref{(3)} and \ref{(4)},
it follows that $\log m \leq c + O(1)$.
Therefore, by Equation~\ref{bound},
\[ d(A_{r}) \leq 2^{c+O(1)}. \]
Here, $c$ is a fixed constant independent of $n$ and $m$.
Since $l(r) \geq l$ and we can choose $l$ arbitrarily,
$d(A_{r}) \leq c_0$ for a fixed constant $c_0$ and
infinitely many $r$, which implies $d(A) \leq c_0$,
and hence the claim.
\end{proof}

We avoided establishing, as in the cited
references, that the elements of $A$ defined
in Equation~\ref{eq.An} are recursive reals.
The resulting proof is simpler, and sufficient for our purpose,
since we only need to establish the finiteness of $A$.

\begin{remark}
\rm
The difficult part of the
Regular KC-Characterization Theorem above
consists in proving that the KC-Regularity Lemma
is exhaustive, i.e., can be used to prove the
nonregularity of all nonregular languages.
Let us look a little more closely at
the set of sequences defined in Item (iii) of the KC-Characterization 
Theorem.
The set of sequences $A$
of Equation~\ref{eq.An} is a superset of the set of characteristic sequences
associated with $L$. According to the proof in the cited 
references, this set $A$ contains finitely many %
\it recursive %
\rm sequences (computable by Turing machines). The subset
of $A$ consisting of the characteristic sequences associated with $L$,
satisfies much more stringent computational requirements, since
it can be computed using only the finite automaton recognizing $L$.
If we replace the plain Kolmogorov complexity in the statement
of the theorem by the so-called `prefix complexity' variant $\KP$,
then the equivalent set of $A$ in Equation~\ref{eq.An} is
\[ \{ \omega \in \{0,1 \}^{\infty} : \forall_{n \in {\cal N}}
[ \KP ( \omega_{1:n} ) \leq  \KP(n) + c ] \} ,\]
which contains nonrecursive
sequences by a result of R.M. Solovay, \cite{So75a}.
\end{remark}

\section*{Acknowledgements.}

We thank Peter van Emde Boas, Theo Jansen,
Tao Jiang for reading the manuscript and commenting on it, 
and the anonymous referees for
extensive comments and suggestions for improvements.
John Tromp improved the proof of Claim~\ref{claim.finite}.

\end{document}